\documentclass[prx,twocolumn,superscriptaddress,citeautoscript,showpacs,amsart,longbibliography]{revtex4-2}
\usepackage{blindtext}
\usepackage{graphicx}
\usepackage{bm}
\usepackage{epsfig}
\usepackage{amssymb}
\usepackage{amsfonts}
\usepackage{braket}
\usepackage{color}
\usepackage{epstopdf}
\usepackage{hyperref}
\usepackage{float}
\restylefloat{table}
\usepackage{bibentry}
\usepackage{color}
\usepackage{multirow}
\usepackage[caption=false]{subfig}
\usepackage{ulem}

\usepackage{amsfonts}
\usepackage{amsthm}
\usepackage{graphicx}
\usepackage{multirow}
\usepackage{color}
\usepackage{bbold}
\usepackage{bm}
\usepackage{times}
\usepackage{amsmath,bm,amsfonts}
\usepackage{dcolumn}
\usepackage{graphicx}
\usepackage{latexsym}
\usepackage{etoolbox}
\patchcmd{\section}
	{\centering}
	{\raggedright}
	{}
	{}
\patchcmd{\subsection}
	{\centering}
	{\raggedright}
	{}
	{}

\newcommand{\bpm}{\begin{pmatrix}}
\newcommand{\epm}{\end{pmatrix}}
\newcommand{\ba}{\begin{eqnarray}}
\newcommand{\ea}{\end{eqnarray}}
\newcommand{\bd}{\begin{displaymath}}

\graphicspath{{figures/}}

\begin{document}
\title{Strain tunable electronic ground states in two-dimensional iridate thin films}

\author{Donghan Kim}
\affiliation{Center for Correlated Electron Systems, Institute for Basic Science, Seoul 08826, Korea}
\affiliation{Department of Physics and Astronomy, Seoul National University, Seoul 08826, Korea}

\author{Byungmin Sohn}
\email[Electronic address:$~~$]{bsohn@skku.edu}
\affiliation{Department of Physics, Sungkyunkwan University, Suwon 16419, Korea}

\author{Yeonjae Lee}
\affiliation{Center for Correlated Electron Systems, Institute for Basic Science, Seoul 08826, Korea}
\affiliation{Department of Physics and Astronomy, Seoul National University, Seoul 08826, Korea}

\author{Jeongkeun Song}
\affiliation{Center for Correlated Electron Systems, Institute for Basic Science, Seoul 08826, Korea}
\affiliation{Department of Physics and Astronomy, Seoul National University, Seoul 08826, Korea}

\author{Mi Kyung Kim}
\affiliation{Department of Physics, Yonsei University, Seoul 03722, Korea}

\author{Minjae Kim}
\email[Electronic address:$~~$]{garix.minjae.kim@gmail.com}
\affiliation{Korea Institute for Advanced Study, Seoul 02455, Korea}

\author{Tae Won Noh}
\affiliation{Center for Correlated Electron Systems, Institute for Basic Science, Seoul 08826, Korea}
\affiliation{Department of Physics and Astronomy, Seoul National University, Seoul 08826, Korea}

\author{Changyoung Kim}
\affiliation{Center for Correlated Electron Systems, Institute for Basic Science, Seoul 08826, Korea}
\affiliation{Department of Physics and Astronomy, Seoul National University, Seoul 08826, Korea}

\date{\today}

\begin{abstract}


Quantum phases of matter such as superconducting, ferromagnetic and Wigner crystal states are often driven by the two-dimensionality (2D) of correlated systems. Meanwhile, spin-orbit coupling (SOC) is a fundamental element leading to nontrivial topology which gives rise to quantum phenomena such as the large anomalous Hall effect and nontrivial superconductivity. However, the search for controllable platforms with both 2D and SOC has been relatively overlooked so far. Here, we control and study the electronic ground states of iridate ultrathin films having both 2D and SOC by angle-resolved photoemission spectroscopy (ARPES) and dynamical mean field theory (DMFT) calculations. The metallicity of SrIrO$_3$ ultrathin films is controlled down to a monolayer by dimensional and strain manipulation. Our results suggest that the iridate ultrathin films can be a controllable 2D SOC platform exhibiting a variety of phenomena for future functional devices.


\end{abstract}
\maketitle

\section*{1. Introduction}

The field of condensed matter physics has advanced remarkably over the past decades, driven by the research on exotic quantum phenomena such as high-temperature superconductivity~\cite{orenstein2000advances,he2023high,chang2012direct}, magnetism~\cite{gibertini2019magnetic}, and metal-to-insulator transition~\cite{imada1998metal,sutter2017hallmarks}. These exotic phenomena often occur due to the two-dimensional (2D) nature of the systems~\cite{cheng2015anomalous}. For example, the cuprate high-temperature superconductivity is known to occur in their 2D square lattices~\cite{orenstein2000advances,halboth2000d,hinkov2004two}. Another prime example is the Kagome lattice in which the interplay between a peculiar 2D lattice structure and electronic wave function leads to flat bands and Dirac cone-like features in the electronic structure~\cite{ye2018massive,kang2020topological}. These nontrivial band features produce exotic phases such as Wigner crystal, ferromagnetic, and superconducting states, which are applicable for future nano-scale devices~\cite{wu2007flat,liu2018giant,neupert2022charge}.

In addition to the reduced dimensionality in 2D, spin-orbit coupling (SOC) is also a quintessential ingredient leading to various non-trivial topological phenomena in materials~\cite{sohn2021sign,sasaki2012odd}. For example, $5d$ transition metal oxides, specifically Sr$_{n+1}$Ir$_n$O$_{3n+1}$, have become of interest due to the large SOC of iridium atoms~\cite{kim2008novel,nelson2022interfacial}. The large SOC gives rise to exotic phenomena including topological Hall effects~\cite{matsuno2016interface,li2019emergent} and interfacial ferromagnetism in various SrIrO$_3$ (SIO) heterostructure systems~\cite{yoo2021large,skoropata2020interfacial,jaiswal2022direct}. In particular, Sr$_2$IrO$_4$ was suggested to exhibit unconventional $J_{\rm eff}$ = $1/2$ superconductivity, attributed to the interplay between the large SOC and a quasi-2D square lattices~\cite{wang2011twisted,kim2014fermi,kim2016observation}. Multiple studies followed to address the conundrum of the nontrivial superconductivity in Sr$_2$IrO$_4$, employing experimental techniques such as pressure-dependent experiments~\cite{haskel2012pressure,haskel2020possible} and chemical doping to control SOC~\cite{zwartsenberg2020spin} and chemical potential~\cite{de2015collapse,nelson2020mott}. In spite of these efforts, many questions remain since the insulating nature of Sr$_2$IrO$_4$ often hamper experimental measurements of material properties~\cite{kim2014fermi,kim2016observation}.

Our alternate approach is to create artificial 2D iridate systems by growing ultrathin iridate films. Growing samples in ultrathin film form provides an additional benefit of controlling the atomic structure of the material by using different substrates and buffer layers. With this motivation, we investigate the electronic structure of ultrathin SrIrO$_3$ (SIO) films down to a monolayer by using angle-resolved photoemission spectroscopy (ARPES)~\cite{sobota2021angle} and dynamical mean field theory (DMFT) calculations. We confirm the metallic ground state of ultrathin SIO films is obtainable through dimension and strain control. Notably, SIO ultrathin films can serve as a special controllable platform as they have both 2D and SOC properties in addition to   nontrivial properties such as $J_{\rm eff}$ = $1/2$ superconductivity or emergent ferromagnetism~\cite{wang2011twisted,kim2014fermi,kim2016observation,jaiswal2022direct,yoo2021large,jaiswal2023giant}. We assert that the SIO ultrathin platform has a potential to be used for future functional devices, $e.g.$, Mott field effect transistors~\cite{shoham2023bandwidth} by utilizing both the dimensionality and inherent SOC of the systems.


\begin{figure*}[htbp]
\includegraphics[width=0.95\textwidth]{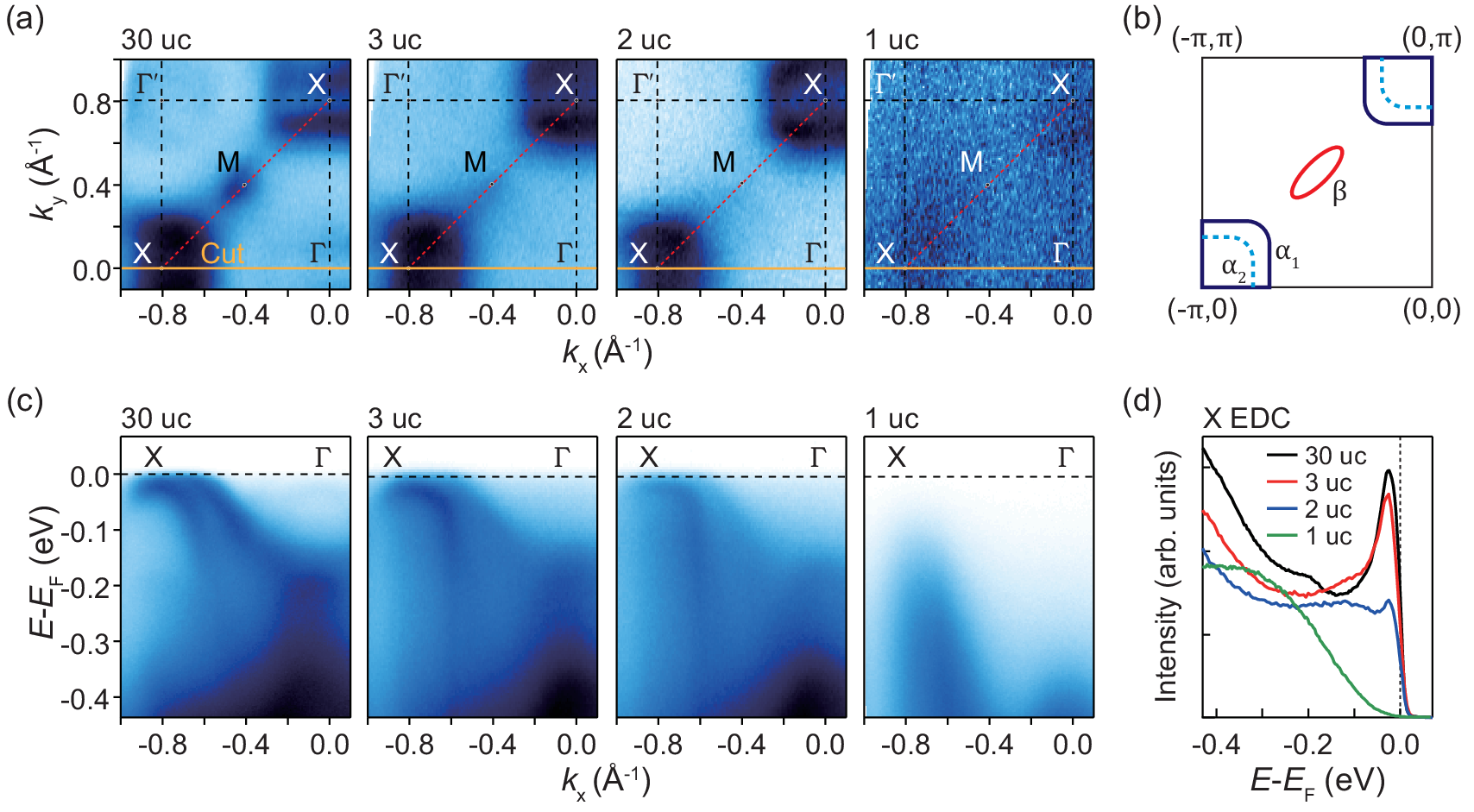}
\caption{(a) Fermi surfaces of SrIrO$_3$ (SIO) films grown on SrTiO$_3$ (STO) substrates as a function of thickness. The folded Brillouin zone by the rotation of IrO$_6$ octahedron is highlighted by a red dashed line (See Fig. S3 for low-energy electron diffraction data). (b) A schematic of the Fermi surface of SIO. $\alpha_1$ and $\alpha_2$ are hole-like bands, and $\beta$ is an electron-like band. (c) Energy-momentum dispersion cuts along the $X$-$\Gamma$ direction. The cuts are represented by orange solid lines in (a). (d) Energy distribution curves (EDCs) at the $X$ point.}
\label{fig:1}
\end{figure*}

\section*{2. Results}
\subsection*{2.1 Thickness-dependent ARPES measurements}
The SIO films with different thicknesses (30, 3, 2, and 1 unit-cell (uc)) are grown on a SrTiO$_3$ (STO) substrate by the pulsed laser deposition method (details of sample preparation and characterization are provided in the Methods section and Fig. S1-2 in the Supplementary Materials). Figure 1 shows the evolution of electronic band structures of SIO films as a function of thin-film thickness. The Fermi surfaces of the SIO films are shown in Fig. 1(a). The 30~uc SIO has two closely spaced hole-like bands, $\alpha_1$ and $\alpha_2$, at the $X$ (($\pm\pi$, 0), (0, $\pm\pi$)) point and an electron-like band, $\beta$, located at the $M$ ($\pm\pi/2$, $\pm\pi/2$) point (see Fig. 1(b) for the Fermi surface schematic)~\cite{nie2015interplay,liu2016direct,liu2021electron}. As the thickness decreases to 2 uc, the electron-like pocket at the $M$ point gradually disappears (Fig. 1(a)), while the spectral weight of two hole-like pockets at the $X$ point remains. Eventually, no spectral weight exists at the Fermi surface in a monolayer SIO.

$\Gamma$-$X$ high symmetry cuts (orange solid lines in Fig. 1(a)) are shown for each thickness of SIO films in Fig. 1(c). The spectral weight near the Fermi level is resolved down to 2~uc, while any band structures do not cross the Fermi level in the 1~uc SIO. Thus, we propose that the 1~uc SIO is insulating on the STO substrate. The valence band top of the 1~uc SIO is located at the $X$ point with $E$ = $E_{\rm F}$ $–$ $0.1$~eV, and the overall band features are similar to those of undoped Sr$_2$IrO$_4$~\cite{de2015collapse}. Note that the $\alpha_1$ and $\alpha_2$ bands are not distinguishable in the 1~uc SIO.

Figure 1(d) shows the energy distribution curves (EDCs) at the $X$ point for different thicknesses of SIO. Sharp quasiparticle peaks~\cite{nie2015interplay} survive down to 2~uc SIO. For 1~uc SIO, however, only an incoherent hump-like peak is observed near $E$ = $E_{\rm F}$ $–$ $0.3$~eV without the quasiparticle peak, reminiscent of the low Hubbard band (LHB) peak in insulating Sr$_2$IrO$_4$~\cite{king2013spectroscopic}. The thickness-dependent changes in EDCs show that a metal-to-insulator transition occurs between the 2~uc and 1~uc SIO films due to the reduction in dimensionality (or, thickness)~\cite{king2014atomic,moon2008dimensionality}.

\begin{figure*}[htbp]
	\includegraphics[width=0.95\textwidth]{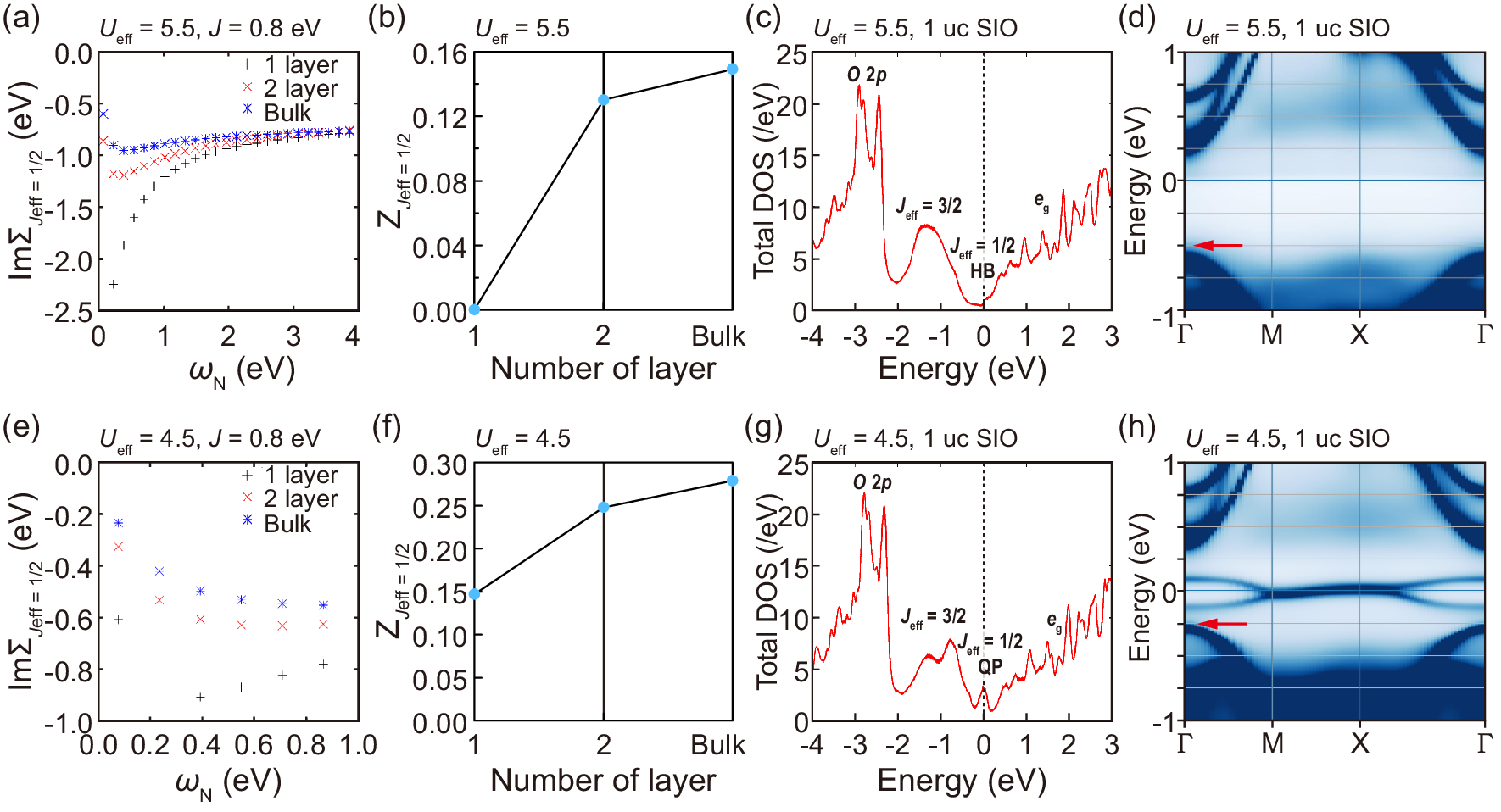}
	\caption{(a, e) Imaginary part of the self-energy (Im$\Sigma$$_{J_{\rm eff} = 1/2}$) in the Matsubara frequency obtained from a dynamical mean field theory (DFMT) calculation for different layers of SIO at effective electron correlation $U_{\rm eff}$ = 5.5 and 4.5, respectively. Exchange interaction J is fixed at 0.8~eV. (b, f) Quasiparticle residue (Z$_{J_{\rm eff} = 1/2}$) of each layer number of SIO at $U_{\rm eff}$ = 5.5 and 4.5, respectively. (c, g) Total density of states (DOS) plot (in the real frequency) of a 1~uc SIO at $U_{\rm eff}$ = 5.5 and 4.5, respectively. (d, h) Spectral function of the 1~uc SIO in the real frequency calculated at $U_{\rm eff}$ = 5.5 and 4.5, respectively.}
	\label{fig:2}
\end{figure*}

\subsection*{2.2 DFT + SOC + DMFT calculations}
To investigate the effect of dimensionality reduction in SIO films, we perform density functional theory (DFT) + SOC + DMFT calculations for ultrathin layers of SIO (see Fig. S4 for DFT calculation results). We define an effective electron correlation (Hubbard interaction) parameter as $U_{\rm eff}$ = $U/W$, where $U$ is the electron correlation and $W$ is the bandwidth, schematically. In the calculations, a bandwidth parameter of $W$ = 1~eV is used as a unit. Figure 2(a) shows the imaginary part of the self-energy in the Matsubara frequency (Im$\Sigma$) for the $J_{\rm eff}$ = $1/2$ state, calculated with $U_{\rm eff}$ = 5.5 and exchange interaction $J$ = 0.8~eV (see Supplementary Note 1 for validity of $U_{\rm eff}$ = 5.5). For two or more SIO layers, a finite value of Im$\Sigma$ is obtained, which corresponds to the metallic state~\cite{bulla2001finite}. However, for a one layer of SIO, as $\omega_{\rm N}$ approaches zero, Im$\Sigma$ diverges to infinity, meaning that no coherent quasiparticles exist and the transition from metal to insulator occurs~\cite{bulla2001finite}. Note that the presence of coherent quasiparticles is associated with a spectral peak near the Fermi level in the photoemission spectra~\cite{damascelli2003angle}. Since the electronic states of SIO near the Fermi level are described as Ir 5$d$ SOC coupled states with $J_{\rm eff}$ = $1/2$~\cite{kim2008novel}, the coherent quasiparticle peak of SIO consists of Ir 5$d$ $J_{\rm eff}$ = $1/2$ states (see Supplementary Note 2 for a detail theoretical description). Related features of the metal-insulator transition can be further verified with quasiparticle residue for $J_{\rm eff}$ = $1/2$ shown in Fig. 2(b). Since only $J_{\rm eff}$ = $1/2$ states can contribute to the density of states (DOS) near the Fermi level in a tetragonal SIO (see Fig. S5 for Green’s function calculation), a zero value for the quasiparticle residue corresponds to an insulating state.

The electronic properties of an insulating 1~uc SIO can be described by the total DOS and the real frequency spectral functions, as shown in Fig. 2(c) and (d). A near-zero value of DOS is obtained at the Fermi level in Fig. 2(c) with no quasiparticle peak. The calculated spectral function of 1~uc SIO also shows no spectral weight near the Fermi level (Fig. 2(d)). These results well explain the insulating band structure experimentally observed in 1~uc SIO (Fig. 1).

We further investigate the ground state of SIO ultrathin films with a relatively low value of $U_{\rm eff}$ = 4.5. As shown in Fig. 2(e), Im$\Sigma$ exhibits finite values in all regions, even for the single layer of SIO. Figure 2(f) also shows that the quasiparticle residue is finite even in the single layer. Figure 2(g) shows the total DOS for a 1~uc SIO with $U_{\rm eff}$ = 4.5. A quasiparticle peak with $J_{\rm eff}$ = $1/2$ states is observed near the Fermi level, indicating that the 1~uc SIO is in the metallic state. In contrast to the insulating band structure in Fig. 2(d), the spectral function of the 1~uc SIO with $U_{\rm eff}$ = 4.5 shows a metallic band structure (Fig. 2(h)). Note that the top of the $J_{\rm eff}$ = $3/2$ band at the $\Gamma$ point (marked with a red arrow) does not cross the Fermi level.

\begin{figure}[htbp]
\includegraphics[width=0.48\textwidth]{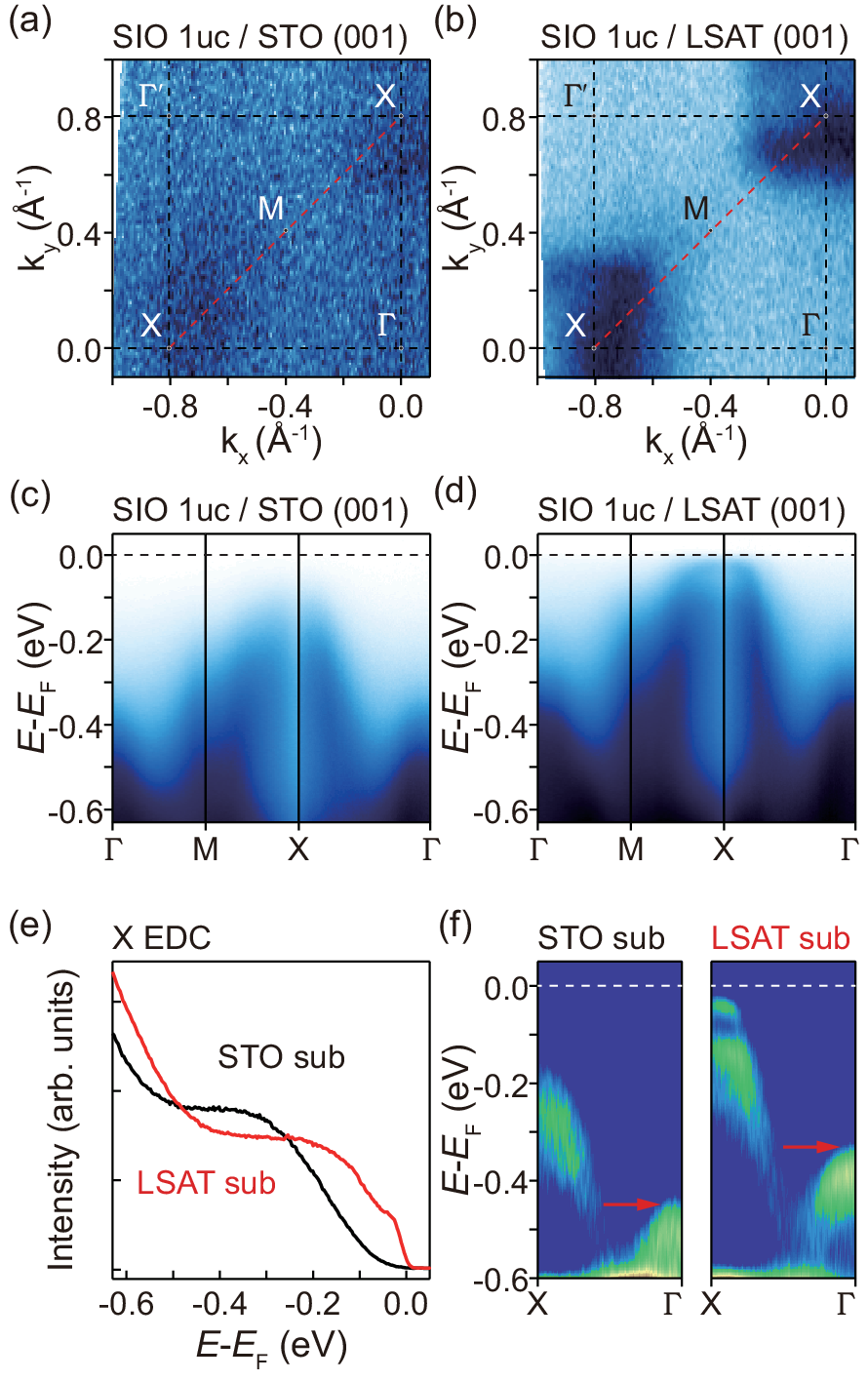}
\caption{(a) Fermi surface of 1~uc SIO on an STO substrate. (b) Fermi surface of 1~uc SIO film grown on an (LaAlO$_3$)$_{0.3}$(Sr$_2$TaAlO$_6$)$_{0.7}$ (LSAT) substrate. (c, d) Energy-momentum dispersion cuts along the $\Gamma$-$M$-$X$-$\Gamma$ direction for 1~uc SIO films grown on STO and LSAT substrate, respectively. (e) EDCs of 1~uc SIO films at the $X$ point. (f) Second derivative plots along energy directions. Spectral weights at the $\Gamma$ point correspond to $J_{\rm eff}$ = $3/2$ state (indicated with red arrows).}
\label{fig:3}
\end{figure}

The DMFT results suggest that the electronic ground state of 1~uc SIO can be manipulated by tuning $U_{\rm eff}$; that is, a small decrease in $U$ or an increase in $W$ can make the insulating monolayer iridate metallic. One possible way to manipulate $U$ or $W$ is to control the strain applied to the film~\cite{shoham2023bandwidth,paris2020strain,rondinelli2011structure}. Several studies have been reported on the effect of strain on the ground state of perovskite iridates, particularly in Sr$_2$IrO$_4$ films~\cite{zhang2013effective,paris2020strain,seo2019compressive} and SIO/STO superlattices~\cite{yang2020strain,kim2017dimensionality}. The complex interplay of changes in Ir-O-Ir bond angles, Ir-O bond lengths, and spin-exchange interactions can lead to a decrease in the bandwidth of perovskite iridates under compressive strain~\cite{zhang2013effective,paris2020strain,seo2019compressive,yang2020strain,kim2017dimensionality}. Such control of $U$ or $W$ via epitaxial strain has been demonstrated in Sr$_2$IrO$_4$ by experimental observations from Raman spectroscopy and resonant inelastic X-ray scattering experiments, and theoretical calculations from DFT and tight-binding models~\cite{seo2019compressive,paris2020strain}. In addition, previous studies on monolayer SrRuO$_3$ films suggest that electronic ground state control is feasible in a monolayer SIO~ \cite{sohn2021observation,kim2023heteroepitaxial,ko2023tuning}.

\subsection*{2.3 Metallic monolayer via strain control}
To control the $U_{\rm eff}$ of 1~uc SIO, we apply more compressive strain by growing the SIO film on a (LaAlO$_3$)$_{0.3}$(Sr$_2$TaAlO$_6$)$_{0.7}$ (LSAT) (001) substrate. The lattice mismatch between SIO and LSAT is $\sim$2.1\%, which is higher than that between SIO and STO ($\sim$1.1\%); note that the lattice constants of SIO, STO, and LSAT are a$_{\rm pc}$ = 3.95~Å, 3.905~Å, and 3.868~Å, respectively. As shown in Fig. 3(a) and (b), the 1~uc SIO on the STO substrate exhibits an insulating Fermi surface, while the 1~uc SIO on the LSAT substrate clearly exhibits a metallic Fermi surface. As predicted by DMFT calculations, a small increase in strain (corresponding to a small increase in bandwidth)~\cite{seo2019compressive,paris2020strain,kim2017dimensionality} in 1~uc SIO induces a transition from the insulating to the metallic state. In short, our results indicate that 1~uc SIO is on the verge of the Mott insulator-to-metal transition, and even a small increase in thickness or strain can lead to the collapse of the Mott state.

Figures 3(c) and (d) show the energy-momentum band dispersion along the high-symmetry lines ($\Gamma$-$M$-$X$-$\Gamma$) of the insulating and metallic 1~uc SIOs, respectively (see Fig. S6 for the ARPES data in the high binding energy region). In Fig. 3(d), the spectral weight is clearly present near the Fermi level at the $X$ point . In Fig. 3(e), the EDCs at the $X$ point show that the metallic 1~uc SIO on the LSAT substrate has both hump-like peak near $E$ = $E_{\rm F}$ $–$ $0.2$~eV and quasi-particle peak near the Fermi level, while the 1~uc SIO on the STO substrate has a clear insulating feature.

To explore the changes in electronic structures with strain, we plot the second derivative of the energy-momentum band dispersion along the $\Gamma$-$X$ direction (Fig. 3(f)). The occupied $J_{\rm eff}$ = $3/2$ state (indicated by red arrows) is observed at the $\Gamma$ points in 1~uc SIO films on LSAT (at $E$ = $E_{\rm F}$ $–$ $0.33$~eV) and STO ($E_{\rm F}$ $–$ $0.45$~eV), showing that the compressive strain causes the $J_{\rm eff}$ = $3/2$ state band to shift to a lower binding energy. These observations are in good agreement with our DMFT results (indicated by red arrows in Fig. 2(d) and (h)).

\begin{figure}[htbp]
\includegraphics[width=0.48\textwidth]{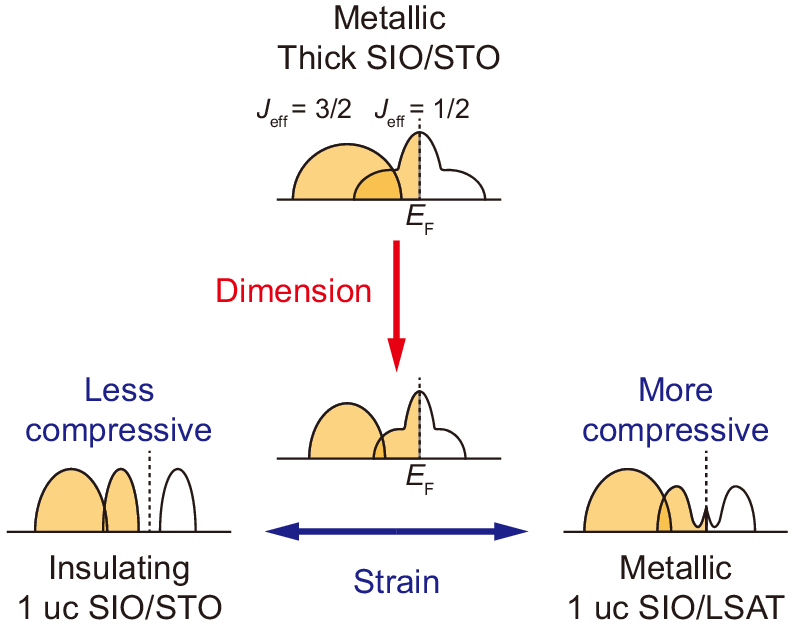}	
\caption{A schematic band diagram of SIO ultrathin films with tunable ground states. Metal-insulator transition can occur depending on dimensionality and applied strain.}
	\label{fig:4}
\end{figure}

In Fig. 4, we present a schematic of our experimental and theoretical findings on the band structure. By reducing the thickness of the SIO film, a bandwidth is reduced in both $J_{\rm eff}$ = $3/2$ and $J_{\rm eff}$ = $1/2$ states, inducing an insulating ground state in the 1~uc SIO film~\cite{groenendijk2017spin}. Note that the critical thickness for the metal-insulator transition is between 2 and 1~uc SIO films, which is confirmed by ARPES and DMFT calculations in this work. In the monolayer limit, the electronic ground state is tunable with strain by substrate engineering. We claim that the compressive strain increases the bandwidth, $W$, of the 1~uc SIO and breaks the Mott insulating state~\cite{seo2019compressive,paris2020strain}.

\section*{3. Discussion}

Note that previous experimental results in SIO films have been reported with a particular focus on the metal-insulator transition. Metal-to-insulator transition of SIO films on the STO substrate has been explored from 3~uc ($\sim$1.2~nm) to 35~nm films~\cite{groenendijk2017spin,schutz2017dimensionality,guo2020engineering,bhat2018influence,biswas2014metal}. In ultrathin films, a critical thickness for the metal-insulator transition is known to be within the range of 3 to 4~uc, as determined through resistivity~\cite{groenendijk2017spin}. Our work shows that the metal-to-insulator transition occurs between 1 and 2~uc in SIO on STO substrates. We propose that the different metal-to-insulator critical thickness is related to the step-edge effect of substrates~\cite{lagally2002thin}. The conducting channel can be broken near the step-edges of substrates when the resistivity is measured, which gives divergent resistivity in mono- and bi-layer SIO films at low temperatures~\cite{groenendijk2017spin}.

In the case of thick SIO films, various properties, including the metal-insulator transition, have been reported~\cite{bhat2018influence,kleindienst2018structural,yang2021generation}. In contrast to the tetragonal structure observed in ultrathin SIO films, the atomic structure of thick SIO films exhibits orthorhombic symmetry~\cite{kleindienst2018structural}. Previous studies on thick SIO films have reported strain-dependent or growth condition-dependent metal-insulator transitions, primarily observed through transport experiments~\cite{bhat2018influence,biswas2014metal}. While direct ARPES measurements could be effective in investigating these observations in thick SIO films, such studies have not been explored yet. Further exploration of ARPES in thick SIO films holds promise for revealing various unexplored phenomena.

\section*{4. Conclusion}

In summary, we control the electronic ground states of SIO ultrathin films by tuning thickness and strain. The Mott insulating state of the SIO is stabilized at a thickness of 1~uc, while the Mott state collapses when small perturbations are introduced into the system. Our DMFT results show that the 1~uc SIO is on the verge of a metal to Mott insulator transition. The newly discovered metallic monolayer state can provide a new platform for studying novel phenomena, such as possible high-temperature superconductivity, which has been intensely investigated in doped Sr$_2$IrO$_4$. Our observations could potentially be used to develop new functional devices in applications. Furthermore, the state tunability of 2D SIO films can help realize a research platform to explore novel emergent phenomena in transition metal oxides, such as $J_{\rm eff}$ = $1/2$ superconductivity or unusual magnetism~\cite{wang2011twisted,kim2014fermi,kim2016observation,jaiswal2022direct}.

\section*{5. Methods}
\subsection*{5.1 Sample preparation}
Epitaxial SIO ultrathin films are grown on SrTiO$_3$ (STO) and (LaAlO$_3$)$_{0.3}$(Sr$_2$TaAlO$_6$)$_{0.7}$ (LSAT) substrates via pulsed laser deposition (KrF; 248~nm wavelength), employing a SrIrO$_3$ ceramic target overdoped with 10~\%\ Ir. Prior to growing SIO ultrathin films, a 10 unit-cell (uc) STO buffer layer is deposited on the substrate, followed by a 10~uc SIO charge reservoir layer, and subsequently, a 10~uc STO buffer layer~\cite{sohn2021observation}. The SIO films are deposited at 600~$^{\circ}{\rm C}$ with an oxygen partial pressure of 100~mTorr, and the STO layers are deposited at 600~$^{\circ}{\rm C}$ with an oxygen partial pressure of 10~mTorr. The energy fluence and repetition rate of the excimer laser for the SIO (STO) layer were 1.5~${\rm J/cm^2}$ (1~${\rm J/cm^2}$) and 2~Hz (1~Hz), respectively. The entire growth process was monitered with {\it in-situ} reflection high-energy electron diffraction (RHEED).

\subsection*{5.2 ARPES measurements}
After film growth, the samples are transferred {\it in-situ} to the preparation chamber and post-annealed at 500~$^{\circ}{\rm C}$ for 10~min. The samples are then transferred {\it in-situ} into the angle-resolved photoemission spectroscopy (ARPES) chamber~\cite{sohn2022evolution,kim2023electric}. Measurements are perfromed with He-I$\alpha$ ($h\lambda= 21.2$~eV) at 10~K using a DA30 analyzer (Scienta Omicron) and a He discharge lamp (Fermi Instruments).

\subsection*{5.3 DFT calculations}
For the DFT band-structure calculations, we employed the full-potential augmented plane-wave band method implemented in the WIEN2k package~\cite{blaha2020wien2k}. We used the local density approximation (LDA) with incorporations of the SOC interaction. We used crystal structures of freestanding monolayer SrIrO$_3$ and bilayer SrIrO$_3$, which have chemical formulas of Sr$_2$IrO$_4$ and Sr$_3$Ir$_2$O$_7$, respectively. The lattice parameter and the rotation angle are adapted from our experimental result for the SrIrO$_3$ thin film on SrTiO$_3$ substrate, a = 3.905~(Å) and c = 4.011~(Å), and the octahedron rotation of 11 degrees in the ${a^0a^0c^+}$ of the Glazer notation. The crystal structure of bulk SrIrO$_3$ is adapted from previous experiments~\cite{zhao2008high}. We used 8$\times$8$\times$1 k-mesh to integrate the Brillouin zone for monolayer and bilayer SrIrO$_3$. For the bulk SrIrO$_3$, we used 16$\times$11$\times$16 k-mesh to integrate the Brillouin zone.

\subsection*{5.4 DFT + SOC + DMFT calculations}
We performed the DFT + SOC + DMFT computation using the TRIQS library~\cite{aichhorn2016triqs,parcollet2015triqs}. The projector is constructed for the whole Ir(5$d$) shell from Kohn-Sham bands in the energy range of [-10,10] eV. The DFT + SOC computation in the LDA is performed using the WIEN2k package~\cite{blaha2020wien2k}. The constructed projector of the Ir(5$d$) shell is transformed to the basis diagonalizes the local field in the local Green’s function from the crystal field and the SOC. This procedure results in the projector of numerical $e_{\rm g}$ and $j_{\rm eff}$ basis. We use this projector in the DMFT computation, taking into account the diagonal part of the local and impurity Green’s function. The Ir($t_{\rm 2g}$) orbital is treated dynamically in the DMFT, while the rest of the valence electrons are treated in the mean-field level. The fully localized limit formalism~\cite{anisimov1997first} is used for the double counting with the nominal occupancy $n_d^0 = 5.0$. We use the density-density type interaction of the Ir(5$d$) electron. The local Coulomb interaction parameters $U$ and $J$ are set as 4.5-5.5~eV and 0.8~eV, respectively. The temperature is set as 290 K. We performed the one-shot DMFT computation. This computational setup for the DMFT is similar to the previous DMFT study on the Ruddlesden-Popper series of strontium iridate~\cite{zhang2013effective}. The paramagnetic state is assumed for whole DMFT computations. The quantum impurity problem in the DMFT was solved using the continuous-time hybridization-expansion quantum Monte Carlo impurity solver as implemented in the TRIQS library~\cite{seth2016triqs,gull2011continuous}.

\vspace{0.5cm}
\acknowledgments
This work was supported by the National Research Foundation of Korea (NRF) grant funded by the Korea government (MSIT) (No. 2022R1A3B1077234) and GRDC (Global Research Development Center) Cooperative Hub Program through NRF (RS-2023-00258359). T.W.N. acknowledges the funding through the Research Center Program of the IBS (Institute for Basic Science) in Korea (grant no. IBS-R009-D1). M.K was supported by Korea Institute for Advanced Study (KIAS) Individual Grants(CG083502). The DFT and the DFT+DMFT calculations are supported by the Center for Advanced Computation at KIAS.

\hfill

\section*{data availability}
The data that support the findings of this study are available from the corresponding author upon reasonable request.


%

\end{document}